\documentclass[preprint,showpacs,preprintnumbers,amsmath,amssymb]{revtex4}

\usepackage{graphicx}
\usepackage{dcolumn}
\usepackage{bm}

\begin{document}
\draft
\preprint{HEP/123-qed}

\title{Itinerant U~5$f$ band states in the layered compound UFeGa$_5$ observed by soft X-ray angle-resolved photoemission spectroscopy}

\author{
Shin-ichi~Fujimori$^1$, Kota~Terai$^1$, Yukiharu~Takeda$^1$, Tetsuo~Okane$^1$, Yuji Saitoh$^1$, Yasuji~Muramatsu$^1$, Atsushi~Fujimori$^{1,2}$, Hiroshi~Yamagami$^{1,3}$, Yoshifumi~Tokiwa$^4$, Shugo~Ikeda$^5$, Tatsuma~D.~Matsuda$^5$, Yoshinori~Haga$^5$, Etsuji~Yamamoto$^5$, and Yoshichika~\=Onuki$^{5,6}$
}
\affiliation{
$^1$Synchrotron Radiation Research Unit, Japan Atomic Energy Agency, Sayo, Hyogo 679-5148, Japan\\
$^2$Department of Complexity Science and Engineering, University of Tokyo, Kashiwa, Chiba 277-8561, Japan\\
$^3$Department of Physics, Faculty of Science, Kyoto Sangyo University, Kyoto 603-8555, Japan\\
$^4$Los Alamos National Laboratory, Condensed Matter and Thermal Physics, MST-10, MS K764, Los Alamos, New Mexico 87545, USA\\
$^5$Advanced Science Research Center, Japan Atomic Energy Agency, Tokai, Ibaraki 319-1195, Japan\\
$^6$Graduate School of Science, Osaka University, Toyonaka, Osaka 560-0043, Japan
}
\date{\today}
\begin{abstract}
We have performed angle-resolved photoemission spectroscopy (ARPES) experiments on paramagnetic UFeGa$_5$ using soft X-ray synchrotron radiation ($h\nu$=500~eV) and derived the bulk- and U~5$f$-sensitive electronic structure of UFeGa$_5$.
Although the agreement between the experimental band structure and the LDA calculation treating U~5$f$ electrons as being itinerant is qualitative, the morphology of the Fermi surface is well explained by the calculation, suggesting that the U~5$f$ states can be essentially understood within the itinerant-electron model.
\\
\end{abstract}

\maketitle
\narrowtext
\section{INTRODUCTION}
Uranium compounds exhibit a rich variety of electrical and magnetic properties, such as heavy Fermion (HF) behaviors, a variety of superconductivity and magnetic ordering, due to their peculiar character of the U~$5f$ states.
The U~5$f$ electrons are located at the boundary between localized and delocalized states, and should be very sensitive to small changes in their environment.
The most fundamental question is how U~5$f$ states are involved in their band structure and Fermi surface (FS), and how they can be described theoretically.
A number of photoemission studies on various uranium compounds have been made so far to clarify this issue.
Especially, angle-resolved photoemission spectroscopy (ARPES) experiments have been made on uranium compounds since the observation of energy dispersions in U~5$f$-derived band is the most direct examination of the U~5$f$ states.
Denlinger {\it et al.} have performed ARPES study on HF superconductor URu$_2$Si$_2$ in the photon energy range of $h\nu =85-156$~eV, including 5$d$-5$f$ resonance ($h\nu = 108$ and $112$~eV) \cite{Allen}.
The obtained band structure and FS were not well explained by the local density approximation (LDA) calculation.
Instead, they have observed the narrow $f$-derived band just below the Fermi level ($E_{\rm F}$), which hybridize with the dispersive non $f$-band.
These features can be understood within the framework of Anderson lattice model treatments involving renormalized hybridization of a renormalized $f$ level to underlying non $f$ states in the vicinity of a crossing of non-$f$ bands \cite{Zwicknagl}.
Therefore, for URu$_2$Si$_2$, the renormalized band theory rather than conventional LDA is more suitable for the description of the U~5$f$ states.
On the other hands, ARPES studies on UAsSe and USb$_2$ in the photon energy range of $h\nu \sim$17.6-60~eV have performed by Los Alamos group, and the good agreement between the experimental results and the LDA calculation was concluded \cite{Arko,LosAramos1}.
They have claimed the itinerant band-like nature of U~5$f$ states in these uranium compounds.
However, in these studies, the ARPES spectra were obtained in very limited part of Brillouin zone, and the agreement with the LDA calculation was not so clear.
Thus, the validity of the LDA calculation for U~5$f$ states in uranium compounds is not still understood, and further investigations are need to clarify this issue. 

Meanwhile, it has recently been recognized that there exist one critical problem in these ARPES measurements.
In these works, the experiments were done with low photon energies ($h\nu =20-120$~eV).
The photoemission experiments performed in this photon energy range are the most surface-sensitive, and may be difficult to relate the observed $f$ band structure to their bulk properties \cite{Bulk}.
Especially, U~5$f$ states should be very sensitive to the surface effects \cite{ULaPdAl}.
Another problem in the ARPES measurements with low-$h\nu$ photons, especially He discharge lamp light, is the very low photoemission cross section of U~5$f$ states compared with those of ligand $s$, $p$, and $d$ states.
The photoemission cross sections of $p$- and $d$- states are more than one order larger than those of $f$-states.
Therefore the obtained energy band dispersions represent contributions mostly from ligand $s$, $p$, and $d$ states, and those from U~5$f$ states were not clearly distinguished\cite{USb,UPt3,URu2Si2}.
In recent years, ARPES experiments in soft X-ray region (SX-ARPES) have been applied to some strongly-correlated electron systems \cite{SXARPES,Suga}.
One of the advantages of this technique is that the spectra are bulk-sensitive compared with conventional ARPES experiments with low-$h\nu$ photons.
For uranium compounds, a further advantage is the large photoemission cross section of U~5$f$ states compared with other ligand $s$, $p$, and $d$ states in the soft X-ray region.
Therefore, by applying this method for uranium compounds, bulk- and U~5$f$ state-sensitive band dispersions and Fermi surface (FS) can be obtained.

In the present study, we have performed ARPES experiments on UFeGa$_5$ using soft X-ray synchrotron radiation ($h\nu $=500~eV) from SPring-8.
We have found that the bulk U~5$f$ electronic structure of UFeGa$_5$ can be obtained by SX-ARPES, demonstrating the usefulness of SX-ARPES experiments for the direct test of LDA calculation for uranium compounds.
Recently, a series of quasi-two-dimensional (2D) $f$-electron compounds with the HoCoGa$_5$-type tetragonal crystal structure have attracted much attention.
Ce-based compounds of this series, Ce$T$In$_5$ ($T$=Rh, Ir, Co) shows interesting competition between magnetism and superconductivity.
For 5$f$ compounds, Sarrao {\it et al.} have discovered that one of this series compound PuCoGa$_5$ undergoes a superconducting transition at $T_{\rm C}$ as high as 18~K \cite{PuCoGa5}.
Also, uranium compounds having this structure show a variety of magnetic properties which are related to the U~5$f$ orbital degree of freedom \cite{U115}.
UFeGa$_5$ is a paramagnetic compound among this series.
dHvA measurement was performed for this compound, and some FS branches were observed \cite{dhva}.
Their topologies were well explained by the LDA band structure calculation including itinerant U~5$f$ states, and the validity of the LDA calculation for 5$f$ states was demonstrated.
The mass enhancement factor is less than 2, suggesting that this compound belongs to one of the most itinerant uranium compounds.
On the other hand, however, it is pointed out that the good agreement between the dHvA signal and the LDA calculation in $f$-based compounds can be viewed as somewhat fortuitous.
In renormalized band theory, mixing of renormalized $f$ band with the non-$f$ bands mainly changes the $f^0$ FS by pushing its $E_{\rm F}$ crossing to accommodate the $f$-electron\cite{Allen}.
This is actually the case of CeIrIn$_5$, where the dHvA signals were well explained by LDA calculation while the ARPES spectra can be interpreted by renormalized band theory\cite{CeIrIn5}.
Thus, it is very important to observed not only the topology of FS but also entire band structure to probe the appropriateness of the LDA calculation for $f$-based compounds.
In addition, it is interesting to see whether the 5$f$ spectra of UFeGa$_5$ show renormalized $f$-band character as in the case of 4$f$ spectra in CeIrIn$_5$, delocalized character as predicted by the LDA calculation, or intermediate between them.
\\
\section{EXPERIMENTAL}
Photoemission experiments were performed at the soft X-rays beamline BL23SU of SPring-8 \cite{BL23SU}.
The overall energy resolution was about 100~meV at $h\nu = 500$~eV.
The sample temperature was kept at 25~K during the course of the measurements.
The position of the Fermi level ($E_{\rm F}$) was carefully determined by measurements of the evaporated gold film.
Single crystals of UFeGa$_5$ were grown by the self-flux method as described in Ref. \cite{dhva}.
Clean sample surfaces were obtained by {\it in situ} cleaving the sample with the surface parallel to the $a$-$b$ plane.
Relativistic-linear-augmented-plane-wave (RLAPW) band structure calculations\cite{Yamagami} within the LDA\cite{LDA} were performed to compare the band structure with the ARPES spectra.

\section{RESULTS AND DISCUSSION}
First, we present the angle-integrated photoemission (AIPES) spectra of UFeGa$_5$.
Figure~1 shows the AIPES spectra of UFeGa$_5$, taken at $h\nu$=400, 500, and 800~eV.
There are two prominent features as designated by arrows.
One is located just below $E_{\rm F}$, and the other is located at about 0.4~eV.
As the photon energy increases, the intensity of the feature just below $E_{\rm F}$ increases, and that of the peak at 0.4~eV decreases.
The relative photoionization cross-section of U~5$f$ states to Fe~3$d$ states is 0.6, 1.0 and 2.5 for $h\nu $=400, 500, and 800~eV, respectively \cite{Atomic}.
Contributions from Ga~4$p$ states are considered to be negligible in these spectra because its cross-section is one order of magnitude smaller than those of U~5$f$ and Fe~3$d$ states in this photon energy range.
Therefore, the feature located at just below $E_{\rm F}$ can be attributed mainly to U~5$f$ states, while the peak structure located at $E_{\rm B}$=0.4~eV is considered to be mainly from the Fe~3$d$ states.
The important point to note is that the spectral intensity in the relatively wide energy region (0.4~eV to $E_{\rm F}$) increases as the incident photon energy increases.
This indicates that the U~$5f$ states are distributed in an energy range of at least 0.4~eV in the occupied side of $E_{\rm F}$.
This behavior is different from those of HF uranium compounds like URu$_2$Si$_2$, where the U~5$f$ states are distributed in a narrower energy region (0.2~eV to $E_{\rm F}$).

To observe the momentum dependence of the spectra, we have measured the SX-ARPES spectra of UFeGa$_5$ at $h\nu $=500~eV.
This photon energy was chosen because the absolute photoemission cross section of the U~5$f$ states has the maximum around this photon energy according to the calculation \cite{Atomic}.
In this photon energy range, the momentum resolution is estimated to be about $\Delta k = \pm 0.02$ \AA$^{-1}$.
Because the size of the Brillouin zone of UFeGa$_5$ is  $\pi / a \sim 0.74$\AA$^{-1}$, about twenty individual $k$ points along one direction in the Brillouin zone can be resolved in the present experiments.
Before we show the ARPES spectra, we evaluate the position of the present ARPES scan in the Brillouin zone.
In this photon energy range, the momentum of the incident photon is not negligible, and the momentum of the electron perpendicular to the surface with a free-electron final state model are given by
%
%
%
%
\begin{equation}
k_{\perp} = \sqrt{\frac{2m}{\hbar^2}(E_{\rm kin}\cos^2{\theta} + V_0)} - k_{\perp{\rm photon}},
\end{equation}
where $E_{\rm kin}$ is the kinetic energy of the photoelectron, $V_0$ is the inner potential, $\theta$ is the emission angle of the photoelectron relative to the surface normal and $k_{\perp{\rm photon}}$ are the momentum of incident photon perpendicular to the surface respectively.
To determine the $k_{\perp}$ value, we assume that the inner potential is 12~eV, which is a typical value for HF compounds \cite{Allen}.
We have evaluated the momentum of incident photon to be $k_{\perp{\rm photon}} = -0.18 {\rm\AA}^{-1}$ in the present experimental setup.
The position of the APRES cut as a function of \mbox{\boldmath $k$} for the $\langle 100 \rangle$ and the $\langle 110 \rangle$ directions is shown as solid line in Fig. 2(a).
Because the escape depth of the photoelectron in the present experiment is about $10 - 15$~\AA, the momentum broadening for the $k_{\perp}$ direction should be $\Delta k_{\perp} \sim 0.06 - 0.1$~${\rm\AA}^{-1}$.
This broadening width ($\Delta k_{\perp} = \pm 0.05$~${\rm\AA}^{-1}$) is also shown as shaded area in Fig. 2(a).
The present ARPES scans trace around the high-symmetry Z-R-A plane with finite $\Delta k_{\perp}$, and therefore we assume that the spectra probe the high-symmetry Z-R-A plane for $k_{\perp}$ direction.

Figure~2 (b) and (c) show the energy distribution curves (EDC's) of UFeGa$_5$ measured along (b) the Z-R (c) the Z-A directions.
Some complicated dispersive features were clearly observed in the spectra for both directions.
For the Z-R direction, some prominent dispersive features were clearly observed.
Especially, in the energy region from 0.6~eV to $E_{\rm F}$, there exist strongly dispersive features centered at the R point.
Clear energy dispersions are observed in the Z-A direction also, and some of them seem to be crossing $E_{\rm F}$.
For example, near the middle of the Z-A direction, one energy band disappears as it approaches to $E_{\rm F}$, and again it appears as a further increase of the momentum, suggesting the existence of the hole-pocket FS along this direction.

To see these features more clearly and to compare them with the results of the band structure calculations, we have derived the ``image'' plot of the band structure by taking the second derivatives of the ARPES spectra.
Overall, the second derivatives of the spectra have been used to reveal positions of the band structure.
In the present procedure, we have further added two processes for this method.
First, we have divided the spectra by the Fermi-Dirac function convoluted with the Gaussian function before making the second derivatives.
This procedure has been used to reveal the structure around $E_{\rm F}$.
Second, we have added the second derivatives of EDC's and MDC's with an appropriate weight
to make the band images.
The former is sensitive to weakly dispersive bands while the later is sensitive to rapidly dispersive bands.
By adding these two contributions, one can make the image plot which reflects both of flat bands and dispersive bands.
We did these procedures very carefully, checking every step so that they do not create spurious structures or wipe out the original weak structures in the ARPES spectra.

Figure~3(a) and (b) shows the experimental band structures for the Z-R and the Z-A directions derived using this method, respectively.
The bright part in the image plot corresponds to the experimental peak position.
For the Z-R direction [Fig.~3(a)], a prominent band is observed in the energy region of 0.6 eV to $E_{\rm F}$.
From its energy position, the feature has the strong contributions from the U~5$f$ states in shallower binding energy sides (0.4~eV to $E_{\rm F}$), and should has the contributions from the Fe~3$d$ states also in higher binding energy sides (0.6~eV to 0.3~eV).
This band forms hole-like FS around the Z point.
The existence of this strongly dispersive energy band suggests that the U~5$f$ states are strongly mixed with the ligand states.
For the Z-A direction [Fig.~3(b)], formation of few FS's are observed.
Particularly, near the middle of the Z-A line, a formation of the hole-pocket, as is seen in the raw ARPES spectra in Fig.~2(c), is well resolved.
These suggest that the U~5$f$ states have large energy dispersions, and form FS in this compound.

We compare these results with the energy band structure calculation.
Figure~3(c) and (d) shows the results of the energy band calculation for the Z-R and Z-A directions, in which the U~5$f$ electrons are treated as being itinerant.
Contribution from the U~5$f$ and Fe~3$d$ states in each band are also indicated on the color scale.
The contribution from the U~5$f$ states is mainly distributed in the energy region from 0.5~eV to $E_{\rm F}$, consistent with the experimental results of AIPES spectra where the contribution from the U~5$f$ states is distributed in the same energy range.
Although the overall agreement between experiment and calculation is not so complete, some calculated bands have a correspondence to the experimental bands.
For the Z-R direction, there exist two or three bands around the Z point in the experimental band structure.
In the band structure calculation, there are few bands in this energy region, and bands 13-15 seem to correspond to the experimental bands among them.
Around the R point, the parabolic band dispersion was observed in the experiment, but it does not exist in the calculation.
Instead, there exist bands 13-15 in the calculation, and the experimental band seems to correspond to the trace of these calculated bands.
These calculated bands have large contributions from U~5$f$ states, suggesting that this experimental bands also should have large contribution from U~5$f$ states.
For the Z-A direction, while the positions of $E_{\rm F}$-crossing are different between the calculation and experiment, the experimental band near $E_{\rm F}$ shows a good correspondence to the calculations.
At the Z point, a hole-like FS is observed in both experiment and calculation.
In addition, the hole-pocket, which is observed in the middle of the Z-A line, also exists in the calculated energy band dispersion.

On the other hand, in renormalized band model, renormalized $f$ bands located near $E_{\rm F}$ are hybridized with non-$f$ bands, and the essential band structure except for the renormalied $f$ bands and the topology of the FS's are very close to those of non-$f$ material. 
To examine the validity of renormalized band model for UFeGa$_5$, we have also shown the result of LDA calculation for ThFeGa$_5$ as dotted lines in Fig. 4(c) and (d).
The essential band structures of UFeGa$_5$ and ThFeGa$_5$ below 0.6 eV, where the contribution from Fe~3$d$ states is dominant, are very similar each other.
However, in the energy range of 0.6 eV to $E_{\rm F}$, there exist strong contributions from U~5$f$ states in the calculated band structure of UFeGa$_5$, and therefore the band structures of the U- and Th-compounds are very different.
In the calculated band structure of UFeGa$_5$, there exist three bands labeled as bands 13-15, which have a large contribution from U~5$f$ states.
Meanwhile, there are few bands in this energy range in the calculated band structure of ThFeGa$_5$ also, but their structures, especially near $E_{\rm F}$ part, are very different.
In particular, although the band 15 shows very good correspondence between the experiment and calculation, the corresponding bands in the calculation on ThFeGa$_5$ cannot reproduce this experimental band.
Comparing these two calculations, it is clear that the agreement with experiment is better for UFeGa$_5$ compared with that of ThFeGa$_5$.
These results indicates that the LDA calculation on UFeGa$_5$ can reproduce the characteristic features of the experimental band structure of UFeGa$_5$, suggesting the itinerant nature of U~5$f$ states in this compound.
In the case of URu$_2$Si$_2$, there is no good correspondence between the experimental bands and the LDA calculation especially near the $E_{\rm F}$ \cite{Allen}.
In other HF compounds also, the renormalized $f$-band behaviors have been observed\cite{UPt3,URu2Si2}.
Therefore, the present result is quite different from the results on other HF uranium compounds.

To obtain more information about the FS's, we have made a 2D mapping of the photoemission intensity at $E_{\rm F}$.
Figure~4(a) shows the mapping as a function of $k_x$ and $k_y$.
The two-dimensional scan was done for the rectangular region including part of the first and the second Brillouin zones.
This area contains the some points with the same symmetry, and therefore they are averaged.
We constructed a mapping of the small triangular region surrounded by the Z-R-A line by this method, and it was eight-fold symmetrize. 
The energy window was chosen to be $E_{\rm F}\pm 50$~meV.
The spectra were divided by the FD function before this procedure to enhance the contribution from just above $E_{\rm F}$.
Bright part corresponds to higher intensity and dark part to lower intensity.
The most prominent feature in this image is the large round-shaped high intensity part centered at the A point, corresponding to the $E_{\rm F}$ crossing in the Z-A direction [Fig. 3(b)].
Its inner and outer boundaries correspond to the hole-like and electron-like FS centered at the A point respectively.
The calculated FS's of UFeGa$_5$ for the Z-R-A plane are also shown in Fig. 4(b) as solid lines.
In the band structure calculation, there exist two large quasi-2D cylindrical FS's centered at the A-M line as indicated in Fig. 4(c).
These cylindrical FS's have large contributions from U~5$f$ states.
Here we note that the shape of the calculated FS is different from the one used in Ref. \cite{dhva}.
In the present calculation, this cylindrical FS is connected to the cross-shaped FS while it is not in Ref. \cite{dhva}.
This connection disappears if the Fermi level is raised by 1-2~mRyd.
Therefore, this feature is very sensitive to the position of the Fermi level, and the difference between these two calculations is subtle.
The comparison of the experimental FS with the calculation shows that although their sizes and shapes are different between the experiment and the calculation, these FS's exist in the experiment also.
Hence, we have obtained a qualitative agreement between experimental FS and the result of the LDA calculation.
We have shown the calculated FS's of ThFeGa$_5$ in Fig. 4(b) as dotted lines.
The two cylindrical FS's centered at A exist in the calculated FS's of ThFeGa$_5$ also, but the calculation predicts more complex FS structure. 
Moreover, the hole-like FS centered at the Z point does not exists in the calculation on ThFeGa$_5$.
Therefore, we conclude that the topology of the experimental FS's is reproduced by the LDA calculation on UFeGa$_5$ rather than ThFeGa$_5$.
This again argues the itinerant nature of U~5$f$ states in this compound.
Meanwhile, an important point to note is that the high intensity part persists from the outer hole-like FS to the inner electron FS as indicated in Fig. 3(b).
This indicates the renormalization of bands around $E_{\rm F}$.
This result shows an existence of electron correlation effect in this compound, and is consistent with the fact that the cyclotron mass of these bands are about two times larger than those of the LDA calculation \cite{dhva}.

Accordingly, we conclude that although the agreement between the calculation and experiment is not complete, the essential band structure and the morphology of FS of UFeGa$_5$ are explained by the LDA calculation.
At the same time, we have observed that the bands near $E_{\rm F}$ are renormalized, suggesting the importance of electron correlation effects even in delocalized uranium compounds.
For the uranium compounds in which the band structures and FS's were studied by ARPES measurements so far, the renormalized band picture has been appropriate for the description of U~5$f$ states rather than the conventional LDA calculation\cite{Allen,USb,UPt3,URu2Si2}.
We believe that UFeGa$_5$ is the first case in $f$-based compounds, where the both of the essential band structure and FS's are well explained by LDA calculation.

Finally, we discuss the comparison with 4$f$-electron compounds. 
In the low-$h\nu$ ARPES study of Ce$T$In$_5$ ($T$=Rh and Ir), similar 2D cylindrical FS's were observed \cite{Ce115}.
However, in these Ce-based compounds, these FS's are formed by ligand $s$ or $p$ states, and the contribution from $f$ states is negligible.
In the case of UFeGa$_5$, contributions from U~5$f$ states are dominant in these FS's, and this is a striking difference between Ce$T$In$_5$ and UFeGa$_5$.
This is due to the more delocalized nature of the 5$f$ states compared with the 4$f$ states.
We observed quite different behavior of $f$ electrons in Ce- and U-based compounds, and therefore different theoretical models are appropriate for their descriptions.

\section{CONCLUSION}
In conclusion, we could successfully observe the bulk- and U~5$f$-sensitive ARPES spectra of UFeGa$_5$, which can be related to the bulk electronic structure of the compound.
We have found the good correspondences of the experimental band structure and FS with the LDA calculation on UFeGa$_5$, but not with that of ThFeGa$_5$. 
This suggests that the LDA is a good starting point for the description of U~5$f$ states in this compound.

\acknowledgments
We would like to acknowledge A.~Yokoya and M.~Takeyama for the improvement of the optical performance of the beam line.
The present work was financially supported by a Grant-in-Aid for Scientific Research from the Ministry of Education, Culture, Sports, Science, and Technology Japan under contact No.15740226, and a REIMEI Research Resources from Japan Atomic Energy Research Institute.


\clearpage

\begin{figure}
\caption{(Online color)Valence-band spectra of UFeGa$_5$ measured with $h\nu $=400, 500, and 800~eV.}
\label{figure1}
\end{figure}
\begin{figure}
\caption{(Online color)(a)Position of ARPES cuts in $k$ space. ARPES spectra of UFeGa$_5$ measured along (b) the Z-R and (c) the Z-A directions. The Brillouin zone is also indicated.}
\label{figure2}
\end{figure}
\begin{figure*}
\caption{(Color)Band structure of UFeGa$_5$ measured along (a) the Z-R and (b) the Z-A directions. Calculated energy band dispersions to be compared with the experiment along (c) the Z-R and (d) the Z-A directions are shown.}
\label{figure3}
\end{figure*}
\begin{figure*}
\caption{(Color)(a)Experimental Fermi surface of UFeGa$_5$ (b)Calculated Fermi surface of UFeGa$_5$ (solid lines.) and ThFeGa$_5$ (dotted lines) in the Z-R-A plane. (c) Calculated three-dimensional Fermi surface of UFeGa$_5$.
}
\label{figure4}
\end{figure*}

\end{document}